\documentclass{ws-procs9x6}

\def\etal {{\it et al.}}

\begin{document}

\title{LORENTZ AND CPT VIOLATION IN THE\\
NEUTRINO SECTOR}

\author{JORGE S.\ D\'IAZ}

\address{Physics Department, Indiana University\\
Bloomington, IN 47405, USA\\
E-mail: jsdiazpo@indiana.edu}

\begin{abstract}
Searches for Lorentz and CPT violation using neutrino oscillations and the prospects for future tests using neutrino time-of-flight measurements and beta-decay experiments are presented.
\end{abstract}

\bodymatter

\section{Introduction}
Lorentz invariance is the symmetry that underlies Special Relativity.
The spontaneous breakdown of this spacetime symmetry can arise in some candidate theories of quantum gravity, such as string theory.\cite{SSB_LV} 
In the search for physics beyond the Standard Model (SM), one path consists of the possibility that some of the observed symmetries of the SM could be broken. 
Following this approach, the Standard-Model Extension (SME) generalizes the SM to incorporate all the possible terms in the action that break Lorentz invariance.\cite{SME} 
The SME is a general framework for Lorentz violation and includes a subset of operators that also break CPT invariance.
The development of the SME triggered a worldwide program searching for relativity violations in all sectors of the SM.\cite{datatables}

\section{Lorentz-violating neutrinos}
In the SME, free neutrinos are described by a Dirac-like equation.\cite{KM2004a}
The study of Lorentz-violating operators of arbitrary dimension\cite{KM2012} allows a direct classification of the observable effects.

\subsection{Neutrino oscillations}
Flavor-mixing operators in the SME have been explored in the construction of realistic models for neutrino oscillations as alternatives to the conventional mass-driven description.\cite{LVmodels}
Additionally, the development of techniques to test Lorenz symmetry with neutrino-oscillation experiments\cite{KM_DKM} has led to several experimental  searches. 
These methods have been used by Double Chooz,\cite{DC_LV1} IceCube,\cite{IceCube_LV} LSND,\cite{LSND_LV} MiniBooNE,\cite{MiniBooNE_LV} MINOS,\cite{MINOS_LV} and Super-Kamiokande.\cite{SK_LV} 
In a recent work, the possibility of neutrino-antineutrino oscillations has been explored for the first time.\cite{RebelMufson} 
Using public data on neutrino interactions in MINOS, the limits on 66 different SME coefficients producing sidereal variations of the event rate were presented at this meeting.
There are also 15 coefficients for Lorentz violation that produce time-independent neutrino-antineutrino oscillations. Preliminary results on the study of these effects using the antineutrino spectrum in the Double Chooz experiment were also presented at the meeting.\cite{DC_LV2}

\subsection{Neutrino time of flight}
The interferometric nature of neutrino oscillations makes them sensitive probes of new physics.
Nevertheless, there exists a set of operators in the SME whose effects are unobservable in oscillations. 
These {\it oscillation-free} operators\cite{KM2012} affect all neutrino flavors in the same way; therefore, their experimental signatures appear in decay processes and group-velocity measurements. 
One method to access these operators is by comparing neutrino speed to that of the photon. Different oscillation experiments are making measurements to determine the neutrino speed. 
In the SME, this quantity can depend on many variables including neutrino energy, direction of propagation, and sidereal time. 
Additionally, SME operators that also break CPT invariance can produce differences between neutrino and antineutrino speeds. 
A general presentation of the theory for time-of-flight measurements can be found in Ref.\ \refcite{KM2012}.

\subsection{Beta decay}
The study of the experimental signatures of oscillation-free operators has led to one particular type of operator that eludes observation in time-of-flight measurements. 
Dimension-three operators introduce momentum-independent modifications of the neutrino energy. 
For this reason, they are absent in the general expression of the neutrino group velocity; hence, there are Lorentz-violating effects that remain completely unexplored to date.
This feature makes beta-decay experiments an interesting probe of spacetime symmetries through the study of these {\it countershaded} relativity violations.\cite{DKL}
The unconventional energy dependence introduced by the SME operators alters the phase space of the antineutrino, leading to modifications of the beta-decay spectrum.
Furthermore, the breakdown of invariance under rotations produces direction-dependent decay rates and sidereal variations.
A general presentation of the theory for beta-decay measurements can be found in Ref.\ \refcite{DKL}.
A summary of the main results is presented below.


For neutron experiments, the simplest Lorentz-violating modification is as an isotropic distortion of the electron count rate. 
The modification is maximal at a well-defined energy, which allows experiments to search for this effect.
The antineutrino-electron angular correlation in the decay of unpolarized neutrons gets modified by anisotropic SME operators.
Unconventional effects in the experimental asymmetry include dependence on the orientation and location of the experiment and sidereal variations. 
Similar effects arise for the asymmetry determining the angular correlation between the antineutrino and the neutron spin in the decay of polarized neutrons.


Tritium decay experiments have been designed for direct measurements of neutrino mass. 
In the presence of Lorentz violation, distortions of the endpoint depend on the neutrino mass but also on the location and orientation of the experiment. Additionally, the endpoint energy can oscillate with sidereal frequency.
Published data from the Mainz and Troitsk experiments implies the first limits on anisotropic effects as well as a tenfold improvement in the limit on the isotropic coefficient, previously constrained using IceCube meson thresholds.\cite{datatables}
These data also allow the study of an effective-dimension-two coefficient that modifies the integrated spectrum near the endpoint energy in the same way as the neutrino mass. 
This coefficient can mimic the mass-squared parameter that controls the shape of the spectrum near the endpoint. 
Since this coefficient can have any sign and also vary with sidereal time, a tachyonic-neutrino\cite{Chodos} behavior can appear.
The first limit on this coefficient is obtained also using published data.


For two-neutrino double beta decay, the simplest Lorentz-violating modification appears as an isotropic distortion of the electron-sum spectrum. 
The unconventional energy dependence introduces a modification at a well defined energy that should guide future experimental searches of this effect.
For neutrinoless double beta decay we find that this decay mode can occur even for massless neutrinos, in which the role of the neutrino mass is replaced by a SME Majorana coupling.

\section{Future outlook}
In recent years, the number of explored experimental signals of the neutrino sector of the SME  has experienced a remarkable boost.
Neutrino oscillations have been the main experimental technique and a considerable section of the coefficient space of the minimal SME has been constrained. 
Nevertheless, many effects remain to be studied.
Additionally, the nonminimal sector offers new effects to be explored.
Furthermore, beam experiments performing measurements of the neutrino speed and beta decay experiments studying neutron, tritium, and double beta decay can now join the worldwide program searching for violations of Lorentz and CPT symmetry.

\section*{Acknowledgments}
This work was supported in part
by the U.S.\ Department of Energy
and by the Indiana University Center for Spacetime Symmetries.

\end{document}